\documentclass[]{interact}
\usepackage{bbm}
\usepackage{threeparttable} 
\usepackage{adjustbox} 
\usepackage{pbox}
\usepackage{comment}
\usepackage{color}
\usepackage{epstopdf}
\usepackage[caption=false]{subfig}

\usepackage[numbers,sort&compress]{natbib}
\bibpunct[, ]{[}{]}{,}{n}{,}{,}

\theoremstyle{plain}

\theoremstyle{definition}

\theoremstyle{remark}

\newcommand{\mymu}{\boldsymbol{\mu}}
\newcommand{\mytheta}{\boldsymbol{\theta}}

\newcommand{\mySigma}{\boldsymbol{\Sigma}}

\newcommand{\myB}{\mathbf{B}}
\newcommand{\myR}{\mathbf{R}}
\newcommand{\myD}{\mathbf{D}}
\newcommand{\myM}{\mathbf{M}}
\newcommand{\myU}{\mathbf{U}}

\newcommand{\myW}{\mathbf{W}}
\newcommand{\myV}{\mathbf{V}}

\newcommand{\myY}{\mathbf{Y}}

\newcommand{\mya}{\mathbf{a}}
\newcommand{\mye}{\mathbf{e}}

\newcommand{\myy}{\mathbf{y}}

\begin{document}

\articletype{REVIEW ARTICLE}

\title{Anomaly Detection in Spatio-Temporal Data: Theory and Application}

\author{
\name{Ji Chen\textsuperscript{a}\thanks{CONTACT Ji Chen. Email: ji.chen@yale.edu}}
\affil{\textsuperscript{a}Center for Outcomes Research and Evaluation, Yale University, New Haven, Connecticut 06510, U.S.A.}
}

\maketitle

\begin{abstract}
This paper provides an overview of three notable approaches for detecting anomalies in spatio-temporal data. The three review methods are selected from the framework of multivariate statistical process control (SPC), scan statistics, and tensor decomposition. For each method, we first demonstrate its technical intricacies and then apply it to a real-world dataset, which is 300 images of solar activities collected by satellite.  Our findings reveal that these methods possess distinct strengths. Specifically, scan statistics excel at identifying clustered anomalies, multivariate SPC is effective in detecting sparse anomalies, and tensor decomposition is adept at identifying anomalies exhibiting desirable patterns, such as temporal circularity. We emphasize the importance of customizing the selection of these methods based on the specific characteristics of the dataset and the analysis objectives.

\end{abstract}


\section{Introduction}
\label{sec: intro}
\begin{figure}[htbp]
	\centering
	\begin{tabular}{ccc}
		\includegraphics[width=0.35 \textwidth]{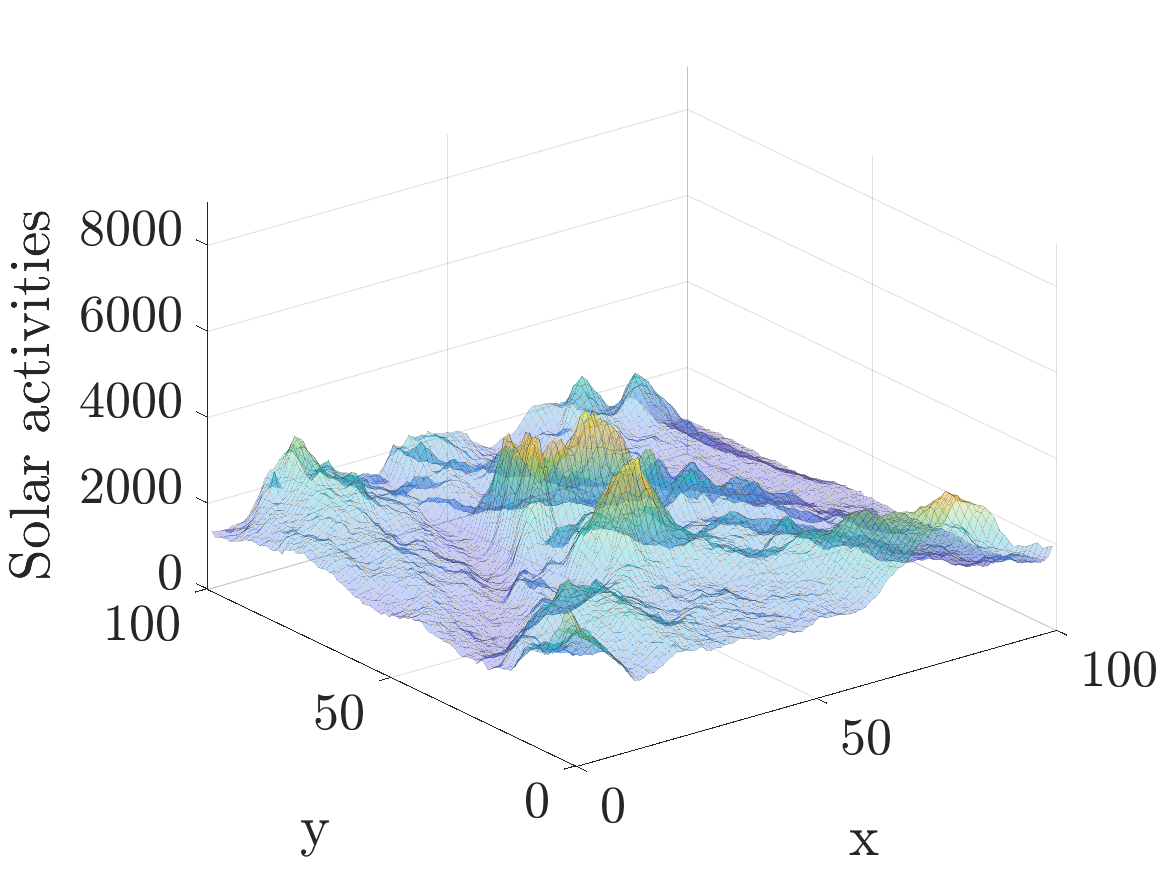} &
		\includegraphics[width=0.35 \textwidth]{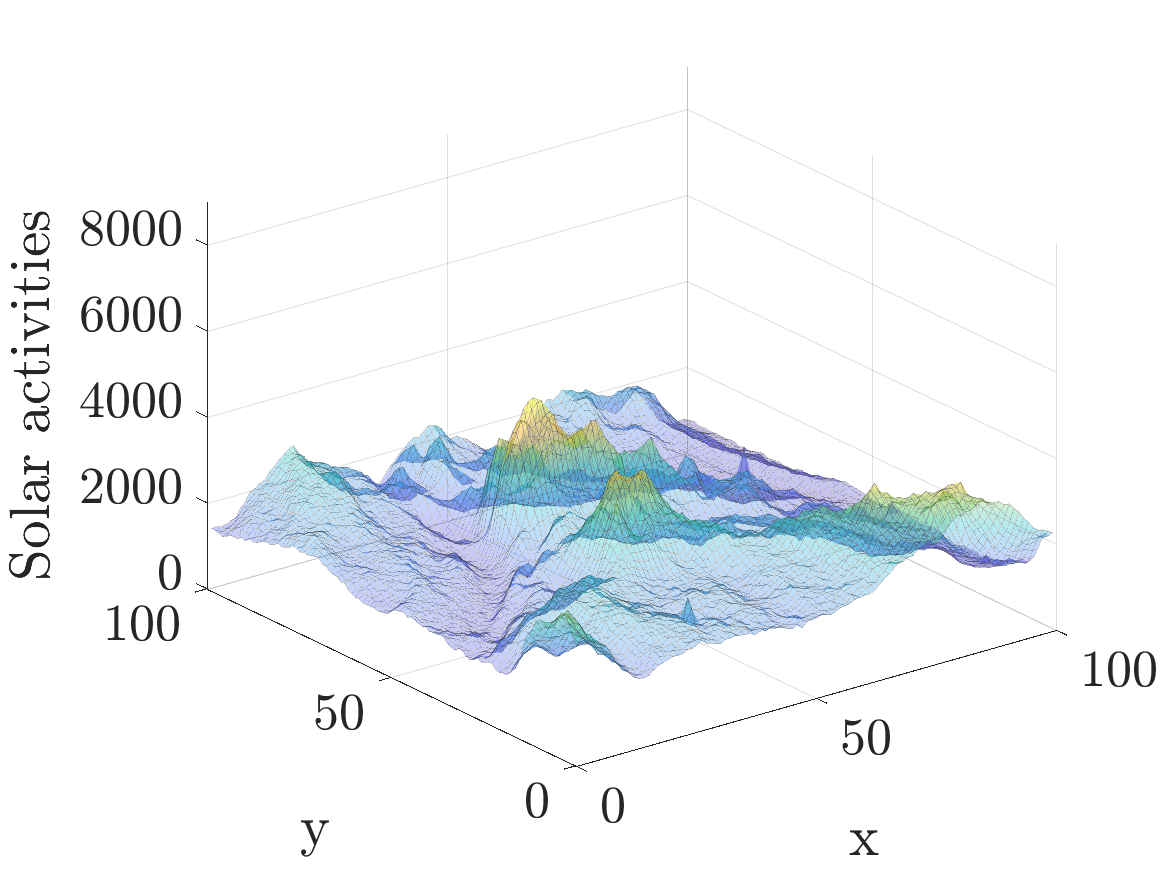}   \\
        (a) $t = 75$ &
		(b) $t = 150$ \\
		\includegraphics[width=0.35 \textwidth]{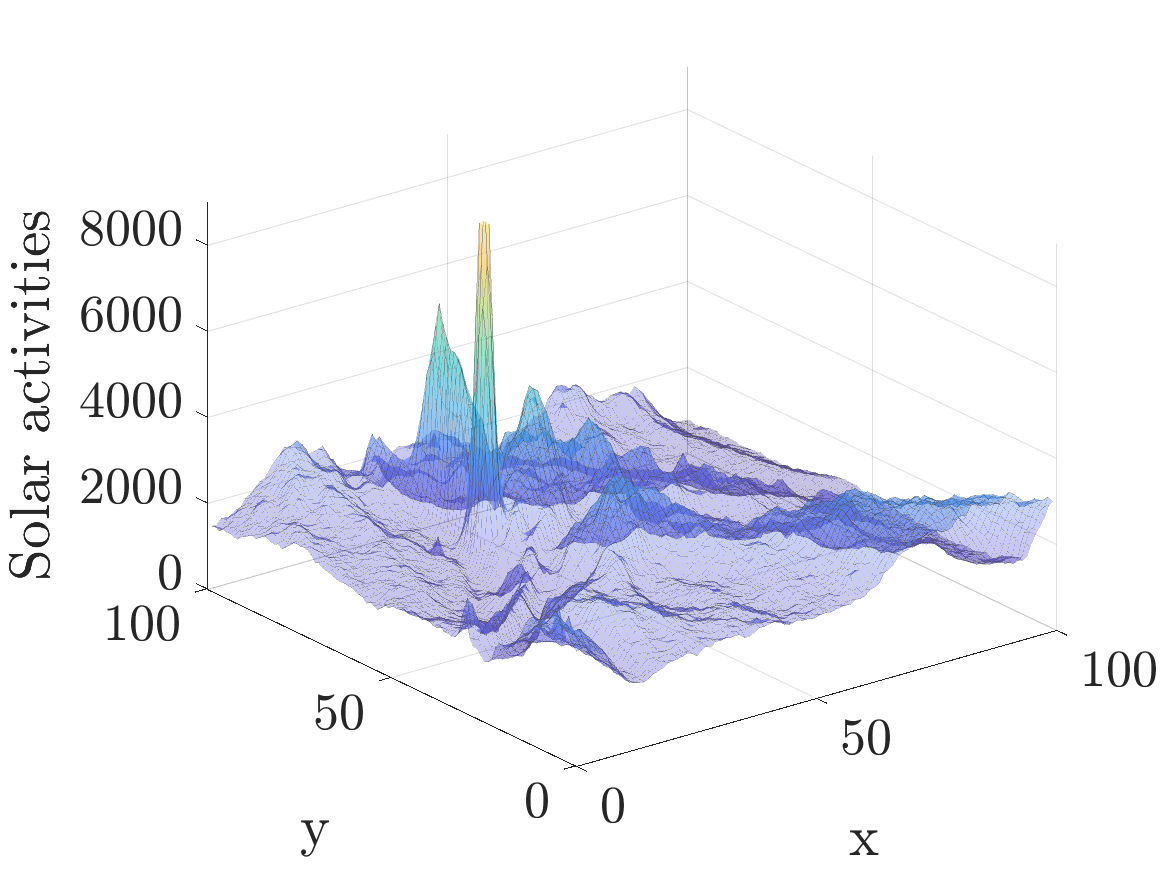} &
        \includegraphics[width=0.35 \textwidth]{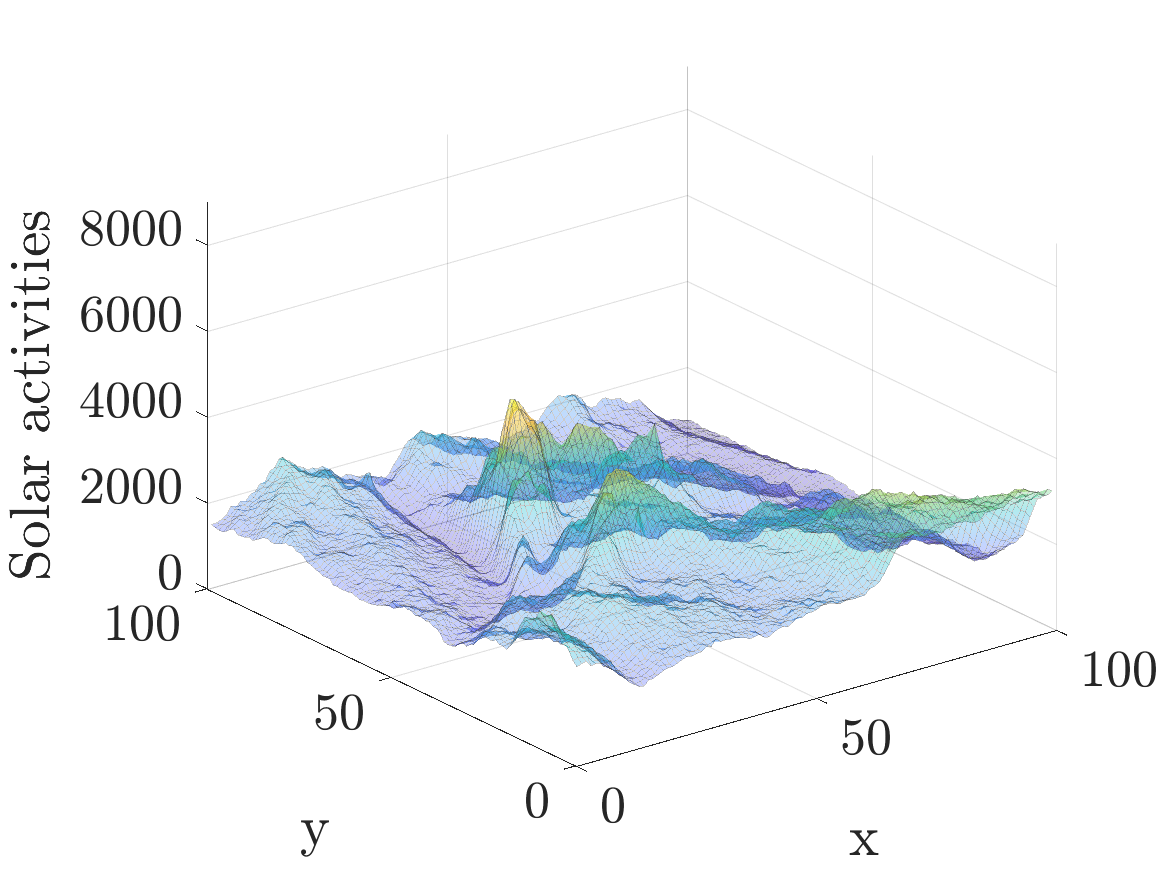} \\
		(c) $t = 225$ &
		(d) $t = 300$ \\
	\end{tabular}
	\caption{
		3D surface plots of the solar activity at time $t = 75, 150, 225, 300$. The x-axis and y-axis are the spatial domains, i.e., $x, y \in \{1, 2, \ldots, 100\}$.
		\label{fig: case study -- surface}}
\end{figure}

Spatial-temporal (ST) data is prevalent in various practical scenarios, such as manufacturing \citep{sergin2021outlier}, epidemiology \citep{zhao2022hot}, and environmental studies \citep{liang2023imputed}. This type of data is typically gathered from multiple spatial locations at regular intervals over time. 
As an example, consider the monitoring of solar activities using solar images consisting of $100 \times 100$ pixels, collected over 300 time points (see a selected presentation of the data in Table \ref{table: solar raw data} and its visualizations in Figure \ref{fig: case study -- surface}). 

Regarding these ST data, there are typically two primary areas of interest: the first one involves modeling fitness and making predictions (see examples in \cite{liang2023imputed, dulal2022covid}), while the second one focuses on outlier detection (see examples in \cite{zou2011multivariate, yan2017anomaly}). In this paper, our emphasis is on the latter, which has its roots in statistical process control (SPC). In the case of one-dimensional or two-dimensional arrays, outliers are commonly referred to as \textit{change points}. However, when it comes to ST data ($d$-way array with $d \geq 3$), the terms \textit{hot-spot} or \textit{anomaly} are more frequently used to denote outliers that occur either in the spatial or temporal domains.

Within the framework of anomaly detection, it is highly valued if we can provide comprehensive information about the identified anomalies. This includes specifying their spatial location (where) and temporal occurrence (when). By offering such detailed information, it becomes convenient for individuals to identify the anomalies and take necessary actions if required.

In the remaining sections of this paper, we will review three representative methods commonly employed for anomaly detection in ST data in Section \ref{sec: method}. Subsequently, we will showcase their performance through an example in Section \ref{sec: application}. Finally, we conclude our paper and offer suggestions for implementers in Section \ref{sec: conclusion}. 

\begin{table}[htbp]
    \caption{A selected presentation of the solar images dataset.}
    \label{table: solar raw data}
    \centering
    \begin{tabular}{cccc}
        \hline
        x-axis & y-axis & Time & Solar activities \\
        \hline
        1      &    1   &  1   & 2968 \\
        2      &    1   &  1   & 2877 \\
        $\vdots$ & $\vdots$ & $\vdots$ & $\vdots$ \\
        100    &    100 & 1    & 862 \\
        1      &    1   &  2   & 1679 \\
        2      &    1   &  2   & 1764 \\
        $\vdots$ & $\vdots$ & $\vdots$ & $\vdots$ \\
        100    &    100 & 2    & 2385 \\
        $\vdots$ & $\vdots$ & $\vdots$ & $\vdots$ \\
        100 & 100 & 300 &  1849\\
        \hline
    \end{tabular}
\end{table}

\section{Methodology}
\label{sec: method}

In this section, we take the tensor of order three $\mathcal{Y} \in \mathbb{R}^{n_1 \times n_2 \times n_3}$ to illustrate the concepts. In the provided example depicted in Figure 1, $n_1 = 100$ corresponds to the number of grids along the x-axis, $n_2 = 100$ represents the number of grids along the y-axis, and $n_3 = 300$ denotes the number of time points. With a fixed time point $t = 1, \ldots, n_3$, the collected data is matrix $\mathcal Y_{::t} \in \mathbb R^{n_1 \times n_2}$, which refers to the image observed at time $t$.

To detect the anomalies in the above ST data $\mathcal Y$, we review three notable methods that are commonly employed in anomaly detection in ST data: multivariate statistical process control (SPC), scan statistics, and tensor decomposition. We select multivariate SPC since anomaly detection is original from the SPC framework. The selection of scan statistics is based on its wide application to detect geographical abnormal clusters. We embrace the tensor decomposition method due to its increasing popularity in recent years for analyzing ST data.

\subsection{Multivariate SPC}
\label{sec: spc}

The field of SPC initially focused on monitoring univariate observations, establishing a solid foundation for control charts such as Shewhart's chart \citep{shewhart1929control}, cumulative sum (CUSUM) \citep{lorden1971procedures, page1954continuous}, exponentially weighted moving average (EWMA) \citep{lowry1992multivariate}, etc. Subsequently, SPC extends its scope from monitoring univariate observations to multivariate observations \citep{zou2009multivariate, liu2019scalable, zou2012lasso}. Within the realm of multivariate SPC, the majority of methods aim to detect anomalies in multivariate data $\myy_1, \myy_2, \ldots, \myy_{n_3}$, where $\myy_t = \text{vector}(\mathcal{Y}_{::t}) \in \mathbb R^{n_1 n_2}$. The primary objective remains the determination of any changes in the mean vector $\boldsymbol{\mu}$ of $\{\myy_t\}_{t=1, \ldots, n_3}$ and identifying the timing of the detected shift in $\boldsymbol{\mu}$. 

To realize the above objective, multivariate SPC assumes the observations $\myy_t$ follow a multivariate normal distribution $N(\mymu_0, \mathbf{\Sigma})$ for $t = 1, \ldots, \tau$, and $N(\mymu_1, \mathbf{\Sigma})$ for $t = \tau + 1, \ldots, n_3$ with known $\mymu_0$ and $\mathbf{\Sigma}$. 
Without loss of generality, let's assume $\mymu_0 = 0$.
To detect the mean shift at time $\tau$, we can test the hypothesis
$$
  H_0: \mymu = 0 \text{  vs.  } H_1: \mymu \neq 0.
$$
Accordingly, the likelihood ratio test statistics at time $t$ is
$$
  n_3 \bar{\mathbf{Y}}_t^\top \mathbf{\Sigma}^\top \bar{\mathbf{Y}}_t
$$
with 
$
  \bar{\mathbf{Y}} = \sum_{i=1}^{t} \mathbf{y}_i / t.
$
If one replaces $\mathbf{\Sigma}$ with the sample covariance matrix, it results in the famous \textit{Hotelling $T^2$ statistics}.  Based on the above test statistics, multiple control charts have been proposed in the framework of the CUSUM or EWMA, where most charting statistics take quadratic forms of the related test \cite{crosier1988multivariate, hawkins2007self}.
With the mean shift signal reported from the control chart, it is possible to employ some diagnostic techniques to determine the specific elements of $\mymu$ that have experienced a shift. Commonly used diagnostic procedures include the step-down procedure mentioned in \cite{mason2002multivariate, sullivan2007step}, as well as the $T^2$ decomposition method described in \cite{mason1995decomposition, li2008causation}.

The aforementioned quadratic testing statistics have demonstrated their effectiveness in control charts, when the objective is to detect shifts occurring in the majority of components in $\mymu$. However, as suggested by \cite{zou2009multivariate, zou2012lasso}, there are some scenarios where the shift in $\mymu$ occurs only in a small number of components with the majority remaining unchanged. In such cases, where the shifts are sparse, it is natural to draw inspiration from the concept of the least absolute shrinkage and selection operator (LASSO) \citep{tibshirani1996regression, zou2012lasso, zhao2021identification}. In \cite{zou2009multivariate}, the estimation of $\mymu$ is motivated by LASSO, where we use a penalized likelihood function
\begin{equation}
\label{equ: spc -- mu hat via lasso}
  \widehat{\mymu}_{t, \lambda}
  =
  \arg\min_{\mymu}
  \left\{
  t(\bar{\myY}_t - \mathbf{\mymu})^\top
  \mySigma^{-1}
  (\bar{\myY}_{t} - \mymu)
  +
  t\lambda 
  \sum_{i=1}^{n_3}
  |\mymu_{t,i}| / |\bar{\myY}_{t,i}|,
  \right\}.
\end{equation}
Here $\mymu_{t,i}$ and $\bar{\mathbf{Y}}_{t,i}$ are the $i$-th element of vector $\mymu_t$ and $\bar{\mathbf{Y}}_t$, respectively. Though the idea of using penalty is the same, we notice the penalty in \eqref{equ: spc -- mu hat via lasso} is slightly different from the traditional one in \cite{tibshirani1996regression}. Specifically, \eqref{equ: spc -- mu hat via lasso} applies different amounts of shrinkage to different regression coefficients, while the regular LASSO applies the same amount of shrinkage for all regression coefficients. With the estimated $\widehat{\mymu}_{t,\lambda}$ from \eqref{equ: spc -- mu hat via lasso}, we derive the LASSO-based test statistics as
\begin{equation}
\label{equ: spc -- lasso test stat}
  T_{t,\lambda}
  =
  \frac{
    t \left(\widehat\mymu_{t,\lambda}^\top \mySigma^{-1} \bar\myY\right)^2
  }{
    \widehat\mymu_{t,\lambda}^\top \mySigma^{-1} \widehat\mymu_{t,\lambda}.
  }
\end{equation}
To calculate \eqref{equ: spc -- lasso test stat}, another crucial block is the selection of the penalty parameter $\lambda$. As high-level guidance, a large value of $\lambda$ is suggested for a large shift, while a small value of $\lambda$ is recommended for a small shift. Specifically, people can use (generalized) cross validation, and some model selection criteria like AIC and BIC. 
In addition to these well-established tools, \cite{zou2009multivariate} propose an alternative strategy, which uses multiple $\lambda$ like $\{\lambda_1, \lambda_2, \ldots, \lambda_q\}$ and finds the most powerful test statistics as
$$
  \widetilde T_t 
  =
  \max_{i = 1, \ldots, q}
  \frac{
    T_{t, \lambda_i} - E(T_{t, \lambda_i})
  }{
    \sqrt{\text{Var}(T_{t, \lambda_i})}
  },
$$
where $E(T_{t, \lambda_i})$ and $\text{Var}(T_{t, \lambda_i})$ are the mean and variance of $T_{t, \lambda_i}$ under the null hypothesis. 
With the aforementioned test statistics, we can build a control chart. Without loss of generality, we take the EWMA as an example and the other control charts like CUSUM can be developed in a similar way. The multivariate EWMA statistics at time $t$ are often defined as 
$
  \mymu_t = \gamma \myy_t + (1-\gamma) \mymu_{t-1},
$
where $\mymu_0 = 0$, $\gamma \in (0,1]$ is a weighting parameter and $\widehat\mymu_t = \mymu_{t, \lambda^*}$ with $\lambda^* =\{\lambda: T_{t,\lambda} = \widetilde T_t \}$. The control chart will raise a signal if 
$$
  \max_{i = 1, \ldots q}
  \frac{
    W_{t, \lambda_i} - E(W_{t, \lambda_i})
  }{
    \sqrt{\text{Var}(W_{t, \lambda_i})}
  }
  > L
$$
where $L > 0$ is a control limit chosen to achieve a given in-control (IC) average run length (ARL), and 
$
  W_{t, \lambda}
  =
  \frac{2-\gamma}{\gamma[1 - (1-\gamma)^{2t}]}
  \frac{
    \left(
    \mymu_t^\top \mySigma^{-1} \widehat \mymu_{\lambda,t}
    \right)^2
  }{
    \widehat \mymu_{\lambda,t}^\top \mySigma^{-1} \widehat \mymu_{\lambda,t}
  }.
$
The first term $\frac{2-\gamma}{\gamma[1 - (1-\gamma)^{2t}]}$, as a conventional practice in the EWMA literature, can be replaced by its asymptotic form $(2-\gamma)/\gamma$. With the time $t^*$ when the signal is raised, we can further localize the anomalies by the non-zero index of vector $\widehat\mymu_{t^*, \lambda^*}$.

\subsection{Scan statistics}
\label{sec: method -- scan stat}
Scan statistics is one of the common statistical tools in cluster detection. When examining a specific time point $t$, it explores spatial regions $S_1, \ldots, S_d$ formed by the observation $\myy_t = \text{vector}(\mathcal Y_{::t})$ to identify regions with notably abnormal measures than expected. To detect these abnormal regions, it tests the possible regions individually and infers if it exhibits significant dissimilarity from the remaining regions. Mathematically, it is developed to test the following hypothesis
\begin{equation}
\label{equ: scan stat --  hypothesis}
  \left\{
  \begin{array}{llc}
     H_0:    & \text{There is no abnormal cluster at time } t    &  \\
     H_1(S): &\text{The region } S \text{ is abnormal at time } t & \forall S \in \{S_1, \ldots, S_d\}. 
  \end{array}
  \right.
\end{equation}
To maintain generality and draw inspiration from the example discussed in Section \ref{sec: intro}, we opt to use the normal distribution presented in \cite{kulldorff2009scan} as an illustrative case, represented by $\myy_{t,i} \sim N(\mu, \sigma^2)$. However, it is important to note that the concept of constructing scan statistics can be applied to alternative distributions, such as the Poisson distribution in \cite{neill2005bayesian} and survival outcome with censoring in  \cite{huang2007spatial, cook2007spatial}.

To test the hypothesis in \eqref{equ: scan stat --  hypothesis}, the scan statistics method finds the abnormal regions $S^*_t$ at time $t$ by maximizing the log-likelihood ratio statistics.
\begin{equation}
\label{equ: scan stat -- test stat}
    S^*_t
    =
    \arg\max_S \{\ln L_S / \ln L_0\}.
\end{equation}
In this way, it detects the region which most likely to be generated under the alternative hypothesis. Here $\ln L_0$ is the log-likelihood under the null hypothesis at time $t$, i.e.,
\begin{equation}
\label{equ: scan stat -- loglikelihood h0}
  \ln L_0
  =
  - n_1 n_2 \ln(\sqrt{2\pi}) - n_1 n_2 \ln(\sigma) 
  - \sum_i^{n_1 n_2} \frac{(\myy_{t,i} - \mu)^2}{2\sigma^2}.
\end{equation}
And $\ln L_S$ is the log-likelihood for region $S$ under the alternative hypothesis at time $t$, i.e.,
\begin{equation}
\label{equ: scan stat -- loglikelihood h1 for S}
  \ln L_S
  =
  - n_1 n_2 \ln(\sqrt{2\pi}) - n_1 n_2 \ln(\sigma_S) 
  - 
  \frac{1}{2\sigma_S^2}
  \left(
  \sum_{i \in S} \left( \myy_{t,i} - \mu_S \right)^2
  +
  \sum_{i \notin S} \left( \myy_{t,i} - \mu_{S^c}\right)^2
  \right),
\end{equation}
where $\mu_S = \sum_{i \in S} \myy_{t,i} / n_S$ is the mean inside of the region $S$ with $n_S$ as the number of observation in region $S$. On the country, $\mu_{S^c}$ is the mean outside of the region $S$. The $\sigma_S$ is the common variance from the maximum likelihood estimation, i.e.,
\begin{equation}
\label{equ: scan stat -- common variance}
  \sigma_S^2
  =
  \frac{1}{n_1 n_2}
  \left(
  \sum_{i \in S} (\myy_{t,i} - \mu_S)^2
  +
  \sum_{i \notin S} (\myy_{t,i} - \mu_{S^c})^2.
  \right)
\end{equation}
By plugging \eqref{equ: scan stat -- loglikelihood h0}, \eqref{equ: scan stat -- loglikelihood h1 for S}, \eqref{equ: scan stat -- common variance} into \eqref{equ: scan stat -- test stat}, we can simplify \eqref{equ: scan stat -- test stat} as
$$
  S^* 
  =
  \arg\max_S 
  \left\{
  n_1 n_2 \ln(\sigma) 
  +
  \sum_{i=1}^{n_1 n_2} \frac{(\myy_{t,i} - \mu)^2}{2\sigma^2}
  -
  \frac{n_1 n_2}{2}
  -
  n_1 n_2 \ln(\sigma_S)
  \right\}.
$$
The above equation aligns with our intuitive understanding: the most likely region selected $S^*$ is the one that minimizes the variance under the alternative hypothesis. 

To determine the statistical significance of the detection region $S^*$, we can use Monte Carlo hypothesis testing \citep{dwass1957modified} with $M$ simulated datasets.  If the log-likelihood ratio from the observed dataset is among the $5\%$
highest of all $M$ datasets, then the most likely region $S^*$ is statistically significant at the 0.05$\alpha$ level. More precisely, the p-value of the region $S^*$ is $R/(M+1)$, where $R$ is the rank of the log-likelihood ratio of the observed dataset in comparison with all $M$ simulated datasets. In practice, suitable choices for $M$ include 999, 4999, or 99999, yielding well-defined p-values with a finite number of decimal places.

\subsection{Tensor decomposition}
\label{sec: tensor}
\begin{figure}[htbp]
	\centering
	\begin{tabular}{ccc}
		\includegraphics[width=0.45 \textwidth]{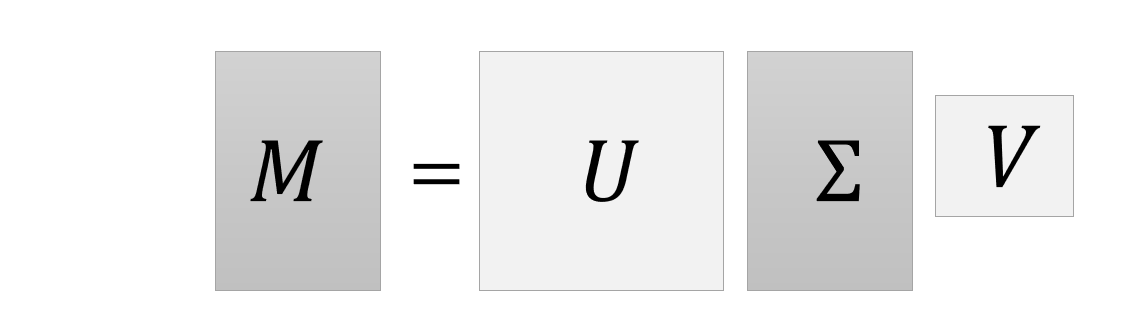} &
		\includegraphics[width=0.45 \textwidth]{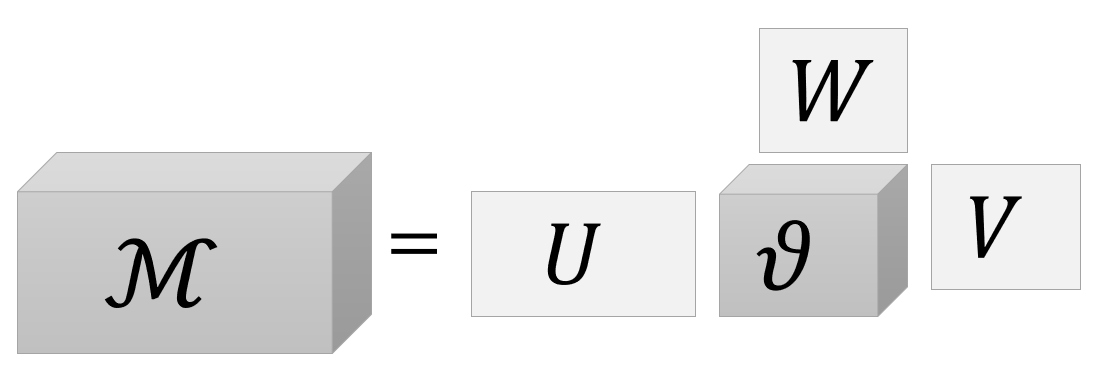}   \\
        (a) SVD &
		(b) tensor decomposition \\
	\end{tabular}
	\caption{
		The relationship between SVD and basis tensor decomposition. In plot (a), we have $\myM \in \mathbb{R}^{n_1 \times n_2}$ as matrix. In plot (b), we have $\mathcal M \in \mathbb{R}^{n_1 \times n_2 \times n_3}$ as tensor of order three.
		\label{fig: method -- tensor -- svd vs tensor decomposition}}
\end{figure}

Tensor data is essentially a $d$-way array $\mathbb R^{n_1 \times n_2 \times \ldots \times n_d}$. And tensor decomposition is a scheme for expressing tensor data as a sequence of elementary operations acting on other, often simpler tensors. Loosely speaking, tensor decomposition is an extension of matrix decomposition and is similar to singular value decomposition (SVD) in the sense of representing a matrix or tensor as the product of several specialized matrices or tensors. Figure \ref{fig: method -- tensor -- svd vs tensor decomposition} illustrates the relationship between SVD and tensor decomposition. From this figure, we can see SVD decomposes a matrix $\myM$ into produce between a core matrix $\mySigma$ and two basis matrices $\myU,  \myV$. Similarly, tensor decomposition decomposes a tensor of order three $\mathcal M$ into a core matrix $\vartheta$ and three basis matrices $\myU, \myV, \myW$. 

In the framework of tensor decomposition, there are multiple options to decompose a tensor, such as rank decomposition in \cite{sidiropoulos2017tensor}, Tucker decomposition in \cite{zhang2016tensor, chuang2009using, dong2010identification}, CP decomposition in \cite{maruhashi2014multiaspectspotting}. While the decomposition scheme may differ, the fundamental concept remains consistent across various decompositions: different tensor products are employed, yet the underlying principle for anomaly detection remains unchanged.

Motivated by the example in Section \ref{sec: intro}, we take the tensor of order three $\mathcal Y \in \mathbb R^{n_1 \times n_2 \times n_3}$ to demonstrate the methodologies in tensor decomposition. Yet, the method can be extended to the tensor of a higher order like $\mathbb R^{n_1 \times n_2 \times \ldots \times n_d}$.  

During the literature review, we discovered that tensor decomposition has been widely utilized for anomaly detection across various domains. These include manufacturing \citep{sergin2021outlier, yan2017anomaly}, health care \citep{zhao2022hot, dulal2022covid}, and social security \citep{zhao2022rapid, opio2022choosing}. 
The focus of most studies has been on identifying local anomalies, which are distinct from the overall group trend. In other words, the interest lies in detecting anomalies with significantly higher values compared to the prevailing global trend. To effectively capture these local anomalies, as proposed by both \cite{yan2017anomaly} and \cite{zhao2022rapid}, it is suggested to decompose the observed data $\myy_1, \myy_2, \ldots, \myy_{n_3}$, where $\myy_t = \text{vector}(\mathcal Y_{::t})$, into three components: a smooth global mean, sparse local anomalies, and random noise:
\begin{equation}
\label{equ: tensor -- model}
    \myy = \mymu + \mya + \mye,
\end{equation}
where $\myy = \text{vector}(\myy_1, \myy_2, \ldots, \myy_{n_3})$. 
By tensor decomposition, we can further decompose the mean $\mymu$ and anomalies $\mya$ as
$$
  \mymu = 
  (\myB_{u,t} \otimes \myB_{u,s}) \mytheta_u,
  \;\;\;
  \mya = 
  (\myB_{a,t} \otimes \myB_{a,s}) \mytheta_a,
$$
where $\otimes$ is the Kronecker product and $\myB_{u,t}, \myB_{u,s}$ are smooth temporal and spatial   bases for the global mean, and $\myB_{a,t}, \myB_{a,s}$ are the temporal and spatial bases for local anomalies. Additionally,
the two vectors $\mytheta_u, \mytheta_a$ are, respectively, the basis coefficients corresponding to $\mymu$ and $\mya$. Specifically, $\mytheta_u = \text{vector}(\mytheta_{u, 1}, \ldots, \mytheta_{u, n_3})$ and $\mytheta_a = \text{vector}(\mytheta_{a,1}, \ldots, \mytheta_{a, n_3})$  where $\mytheta_{u,t}$ and $\mytheta_{a,t}$ are the coefficients of the global mean and local anomalies at time $t$, respectively. We assume that noise $\mye$ is normally independently distributed with mean zero and variance $\sigma^2$.
To estimate these two coefficients (i.e., $\mytheta_u, \mytheta_a$), the penalized least square regression is used:
\begin{equation}
\label{equ: tensor -- objective function}
  \arg\min_{\mytheta_u, \mytheta_a} 
  \|\mye\|^2 + \mathcal P(\mytheta_u, \mytheta_a),
\end{equation}
where $\|\cdot \|$ is the $\ell_2$ norm operators. For the selection of the penalty term $\mathcal P(\mytheta_u, \mytheta_a)$, there are multiple options in the existing literature. For example, \cite{yan2018real} design the penalty to ensure the smoothness of the estimated mean and the sparsity of the detected anomalies, so they set $\mathcal P(\mytheta_u, \mytheta_a)$ as
    $
      \mathcal P(\mytheta_u, \mytheta_a) = \mytheta_u^\top \myR \mytheta_u + \gamma \|\mytheta_a \|_1,
    $
    where $\gamma$ are tuning parameters to be determined by the user. The matrix $\myR$ is a regularization matrix that controls the smoothness of the global mean $\mymu$, and the $\ell_1$ penalty encourages the sparsity of the anomalies. Following \cite{xiao2013fast}, we can set $\myR = \myR_t \otimes \myB_{u,s}^\top \myB_{u,s} + \myB_{u,t}^\top \myB_{u,t} \otimes \myR_s + \myR_t \otimes \myR_s$, where the matrix $\myR_s$ and $\myR_t$ are the regularization matrices that control the smoothness in the spatial and temporal directions. In the motivating example shown in Section \ref{sec: intro}, where the observed data is a sequence of image data, we can set $\myR = \myR_{sx} \otimes \myR_{sy}$ where $\myR_{sx} = \lambda_{s,x}\myD_{s,x}^\top \myD_{s,x}$ and $\myR_{s,y} = \lambda_{s,y}\myD_{s,y}^\top \myD_{s,y}$ are regularization matrices in the x-axis and y-axis, respectively. If smoothness of the global mean is directly related to the difference between two neighbor coefficients, we can set $\myD_{s,x}$ and $\myD_{s,y}$  as first-order difference matrix $\myD_{s,x} = (d_{i,j}) = \mathbbm 1\{j = i\} - \mathbbm 1\{j = i+1\}$ with $\mathbbm 1\{\cdot\}$ as an indicator function. Different from \cite{yan2018real}, \cite{zhao2020rapid} and \cite{zhao2021rapid} aim to detect anomalies with both sparsity and temporal continuity, so they set $\mathcal P(\mytheta_u, \mytheta_a)=
      \lambda \|\mytheta_a \|_1 + \gamma \| \myD \mytheta_a \|_1,
    $
    where $\myD = \myD_{s} \otimes \myD_t$ , matrix $\myD_{s}$ as identical matrix, and 
    $$
      \myD_t
      =
      \left(
      \begin{array}{cccc}
          1 & -1 & &\\
            & \ddots & \ddots &\\
          -1&        &    &   1
      \end{array}
      \right).
    $$
    In this way, the estimated $\mytheta_a$ coefficient has a circular pattern in the temporal domain. In conclusion, we find there are various ways to define penalty, depending on the desirable patterns of $\mytheta_u$ and $\mytheta_a$, and the propriety of the dataset $\mathcal Y$.

To apply the aforementioned tensor decomposition model for anomaly detection, we need to consider several blocks.

The first block involves the computation of \eqref{equ: tensor -- objective function}. As recommended by both \cite{yan2017anomaly, zhao2022hot}, it is possible to adopt established optimization techniques from the field of optimization research, such as the block coordinate descent method \citep{yan2017anomaly, zhao2022survey}, proximal gradient method \citep{yan2017anomaly}, or the fast iterative shrinkage threshold algorithm \cite{zhao2022hot} to solve \eqref{equ: tensor -- objective function} efficiently.

The second block is the selection of the tuning parameters. A common approach is to use the $k$-fold cross-validation and find the parameters that minimize the mean squared error \cite{zhao2020rapid,yan2017anomaly}.

The third block is the basis selection for $\mathcal{B}_u$ and $\mathcal{B}_a$. Typically, when the background exhibits smoothness, it is advisable to utilize a smooth basis, such as splines or kernels. 
If the anomalies are small regions scattered across the background or appear as thin lines, it is recommended to adopt an identity basis. However, if the anomalies form clustered regions, a spline basis may be a more suitable choice. Choosing the anomaly basis is often challenging, but incorporating prior information about the size of anomalies, if available, can aid in the selection process.

The final block involves the construction of control charts. At each time $t$, we test whether the expected residuals after removing the global mean $\mymu_t$ is zero
or has a mean shift in the direction of estimated anomalies $\widehat\mya_t$.
To test these hypotheses, a likelihood ratio test is applied
to the residuals at each time $t$, that is, $\widetilde\mye_t = \myy_t - \widehat\mymu_t$.
This leads to the test statistic 
$$
  T_{\gamma}(t) = (\widehat\mya_t^\top \widetilde\mye_t)^2 / \widehat\mya_t^\top \widehat\mya_t
$$
in which it is assumed that the residuals $\widehat\mye_t$ are independent after removing the global mean and their distribution before and after the change remains the same. Considering the test statistics $T_{\gamma}(t)$ depends on the turning parameter $\gamma$, one can apply the strategy reviewed in Section \ref{sec: spc} to get the most powerful statistics among a pool of $\gamma$. With the selected most powerful test statistics, we can employ it in a control chart. \cite{yan2017anomaly} use the Shewhart control chart, while \cite{zhao2021rapid} utilize the Cumulative Sum (CUSUM) control chart. There is no definitive choice as each control chart has its advantages and suitability depends on the specific type of anomalies under consideration. For instance, if the focus is on detecting anomalies with temporal continuity, the CUSUM control chart is a suitable option. In practice, the selection of the appropriate control chart should be tailored to the specific characteristics of the anomalies of interest. With a built control chart, one can decide the control limit by following the procedure listed below. We first estimate the global mean $\mymu$ and local anomaly $\mya$ from the in-control sample $\myy_1, \myy_2, \ldots \myy_{\tau}$. Then we can determine the control limit as a certain quantile of the empirical distribution of the monitoring statistics based on a predetermined IC average run length. If the control chart raises alarm at time $t^*$, then the detected anomalies are declared to be located at $\{(x,y) : \mathcal Y_{ijt^*} \neq 0\}$.

\section{Applications}
\label{sec: application}

In this section, we employ the aforementioned methods to analyze a real-world dataset, aiming to compare their performance. The dataset consists of a stream of 300 solar images, each measuring $100 \times 100$ pixels, obtained from a satellite. Figure \ref{fig: case study -- surface} presents visualizations of the 75th, 150th, 225th, and 300th frames from the dataset. These images are utilized for solar activity monitoring and the detection of solar flares. Given that solar flares can release a substantial number of highly energetic particles, which have the potential to disrupt large-scale power grids, it is crucial to swiftly identify and respond to solar flare occurrences. Notably, as shown in Figure \ref{fig: case study -- surface}, we observe distinctive behavior in the 225th frame, exhibiting a peak with values surpassing 7000. Consequently, we guess there might be a potential solar flare outbreak at $t = 225$ within the region $R$. To quantify the anomaly, we apply the three aforementioned methods. For analysis purposes, we designate the first 100 frames as in-control sample (anomalies-free).

The implementation details for the three approaches are outlined as follows. For the multivariate SPC method, we specify the values of $\{\lambda_1, \lambda_2, \ldots, \lambda_q\}$ as $\{0.01, 1, 2, 3, 5, 9, 11, 30, 50, 90, 120, 200, 240, 280, 320, 550, 1000\}$. In the tensor decomposition approach, we adopt a B-Spline basis with 50 knots, denoted as $\myB_{u,x}$ and $\myB_{u,y}$, to model the smooth functional mean. To represent sparse anomalies, we utilize the identity matrix as $\myB_{a,x}$ and $\myB_{a,y}$. To ensure a smooth global mean and sparse local anomalies, we follow the approach presented in \cite{yan2017anomaly} and set $\lambda_{s,x} = \lambda_{s,y} = 0.1$ for the $x$ and $y$ axes, enforcing background smoothness. Furthermore, we employ the same penalty parameter as \cite{yan2018real} and set the value of the penalty parameter $\gamma$ equal to $\{\lambda_1, \lambda_2, \ldots, \lambda_q\}$ used in the multivariate SPC method.

\begin{figure}[htbp]
	\centering
	\begin{tabular}{ccc}
		\includegraphics[width=0.3 \textwidth]{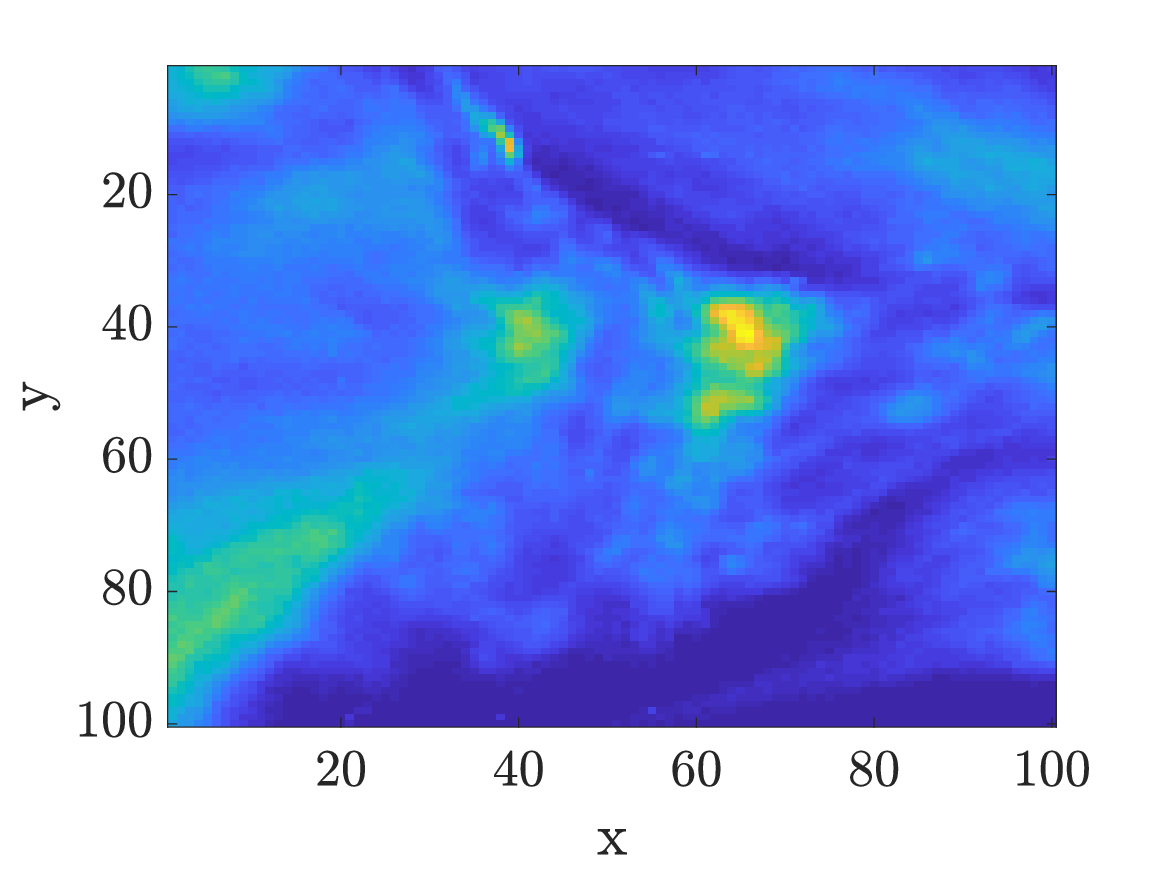} &
		\includegraphics[width=0.3 \textwidth]{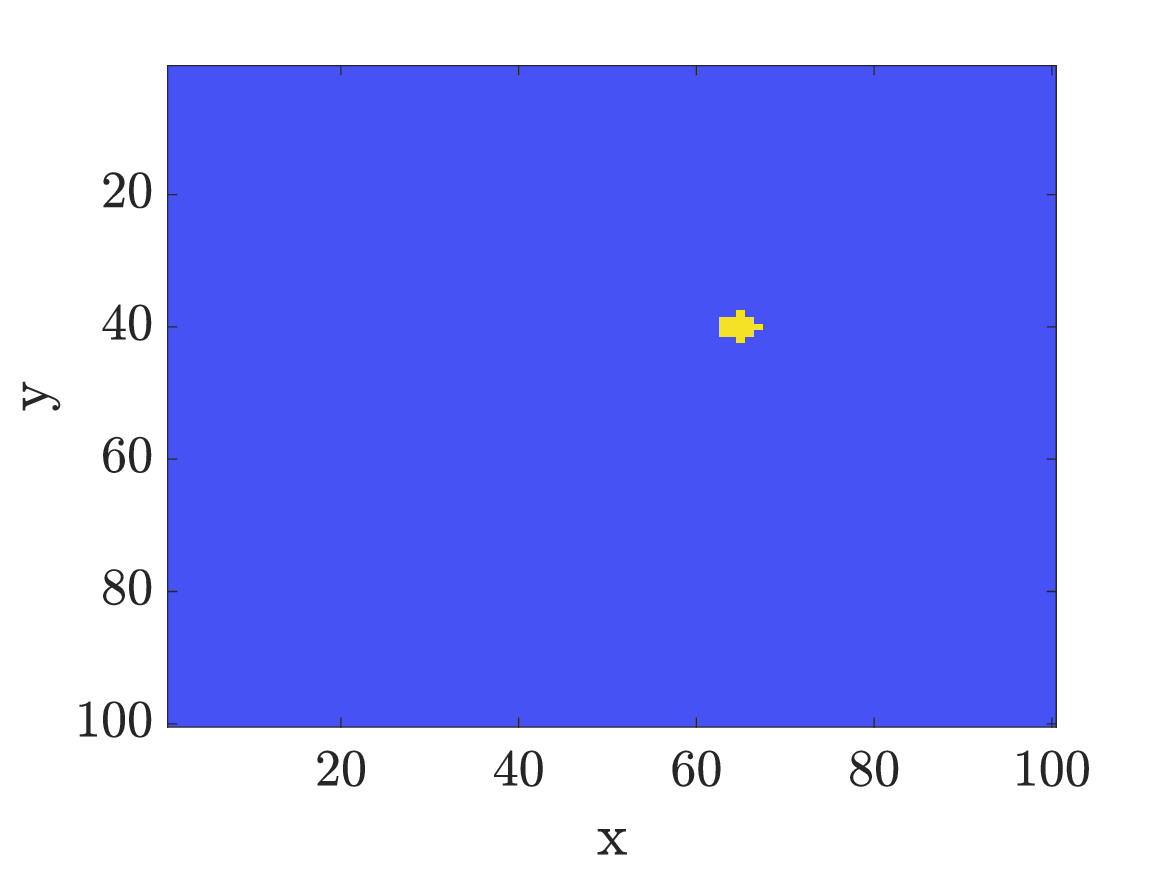}&
        \includegraphics[width=0.3 \textwidth]{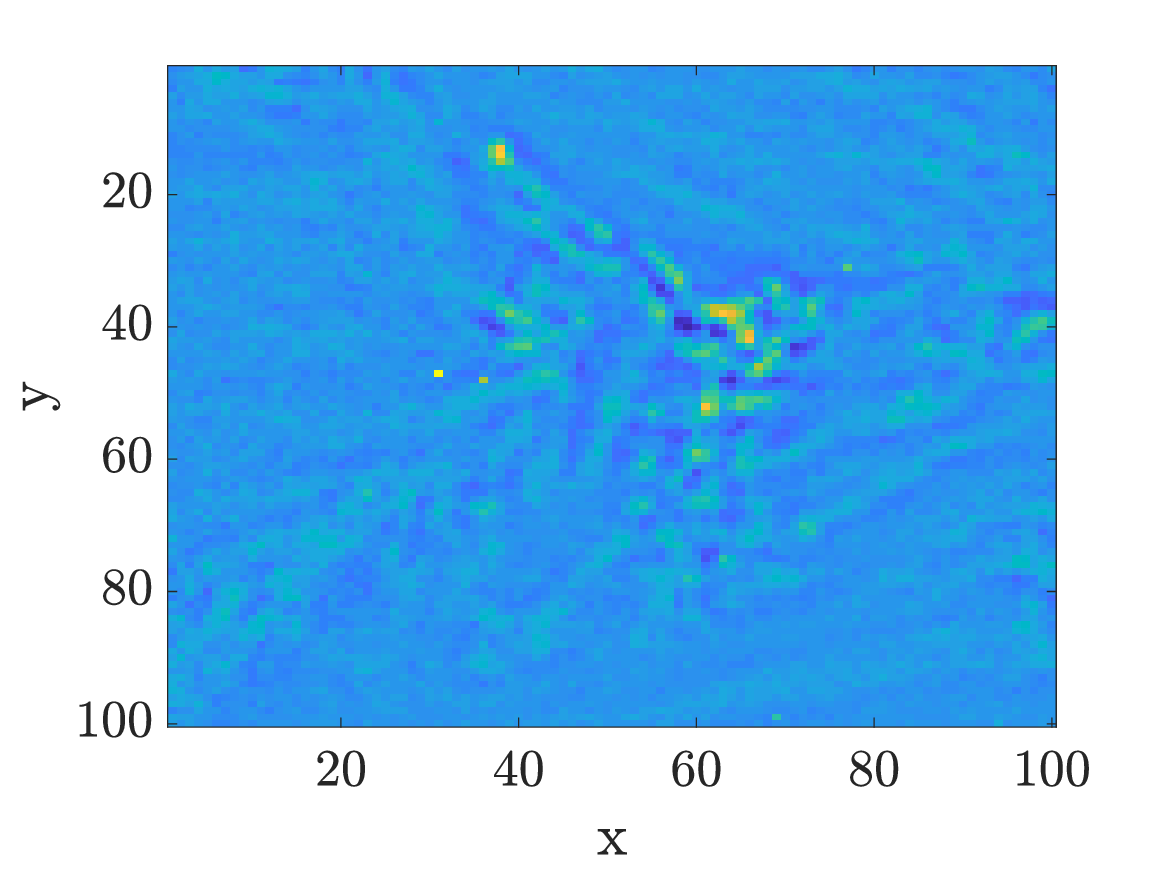}\\
        (a) multivariate SPC &
		(b) scan statistics &
        (c) tensor decomposition 
	\end{tabular}
	\caption{
		Anomalies localized by the three methods reviewed in Section \ref{sec: method}. The yellow color represents the anomalies.
		\label{fig: case study -- anomaly}}
\end{figure}

The summary of the detection results is provided below. The multivariate SPC method detects anomalies in the 195th frame, and its specific location is highlighted as yellow in Figure \ref{fig: case study -- anomaly}(a). Using scan statistics, we localize anomalies as shown in Figure \ref{fig: case study -- anomaly}(b). It is important to note that scan statistics typically indicate the location of anomalies without specifying their values, so we use two different colors to discriminate the anomalies (yellow) and non-anomalies (blue). The tensor decomposition method detects anomalies occurring in the 188th frame, and their locations are visualized in Figure \ref{fig: case study -- anomaly}(c).

By comparing these three methods, we have three interesting findings. First, all three methods detect anomalies at the area of $\{(x,y): 60 \leq x \leq 80, 35 \leq y \leq 45\}$. Second, scan statistic tentatively detects cluster anomalies. This is because the scan statistic tests region by region. Third, the multivariate SPC method performs very similarly to the tensor decomposition. Specifically, they both report the anomalies to occur around the 190th frame. Additionally, the localized anomalies share very similar positions. Yet, these two methods differ in the aspect that, multivariate SPC tentatively detects the global anomalies (since it failed to separate the global mean from the local anomalies), while the tensor decomposition targets local anomalies, which is the one has higher/lower values than the global mean. 

Upon conducting the comparison, we conclude that there is no definitive golden standard method that universally outperforms the others. Instead, the selection of the appropriate method should be tailored to the specific characteristics of the dataset under consideration. For instance, if the focus lies on identifying clustered anomalies, the scan statistics method proves to be a suitable option. On the other hand, if the objective is to detect sparse anomalies, employing $\ell_1$ penalties, as demonstrated in equation \eqref{equ: spc -- mu hat via lasso} of the multivariate SPC method and equation \eqref{equ: tensor -- objective function} of the tensor decomposition method, is recommended. Moreover, if the goal is to detect anomalies relative to the global mean, then the tensor decomposition method emerges as a favorable choice.

\section{Conclusion}
\label{sec: conclusion}

This paper provides an overview of three notable approaches for detecting anomalies in multivariate ST data. For each method, we demonstrate its technical intricacies and its application on a real-world dataset. Our findings reveal that these methods possess distinct strengths (see a summary in Table \ref{table: pros and cons}). Specifically, scan statistics excel at identifying clustered anomalies, multivariate SPC is effective in detecting sparse anomalies, and tensor decomposition is adept at identifying anomalies exhibiting desirable patterns, such as temporal circularity. We emphasize the importance of customizing the selection of these methods based on the specific characteristics of the dataset and the analysis objectives.

\begin{table}[htbp]
  \caption{Pros and cons of the three reviewed methods for anomalies detection 
  \label{table: pros and cons}}
  \centering
  \begin{adjustbox}{max width=1\textwidth}
  \begin{threeparttable}
  \begin{tabular}{p{0.16\textwidth}|p{0.42\textwidth} | p{0.42\textwidth}}
    \hline
    & Pros & Cons \\
    \hline
    \pbox{5cm}{Multivariate \\ SPC} & 
    \pbox{20cm}{ 
      It is a direct extension of univariate \\ SPC, which is a widely used \\ technique for change-point detection. \\ In the context of anomaly detection, \\ it is capable of effectively identifying \\ sparse anomalies.
    }\vspace{5pt}
    & 
    \pbox{20cm}{ 
      It blends anomalies with global mean.\\ If there is an increasing global trend, \\ it tentatively reports most observations \\ as anomalies.  
    } \\
    \hline
    \pbox{5cm}{Scan \\ statistics} & 
    \pbox{20cm}{ 
      It naturally stems from the principle \\ of hypothesis testing. In the context \\ of anomaly detection, it can detect \\ clustered anomalies.
    }
    & 
    \pbox{20cm}{ 
      The computation of the p-values can \\ be time-consuming and it has limited \\ ability to report when the anomaly \\ happens.}
    \vspace{5pt}
    \\
    \hline
    \pbox{5cm}{Tensor \\ decomposition} & 
    \pbox{20cm}{ 
      In the context of anomaly detection, \\ it can be applied to $d$-way array \\ ($d \geq 3$), and detect anomalies with \\ desirable patterns (e.g., spatial \\ sparsity, temporal circularity) which \\deviate from the global mean.
    }
    &
    \pbox{20cm}{ 
      It requires users to be familiar with \\ tensor algebra and the design of the \\ optimization algorithm is not trivial.
    }
    \vspace{5pt}
    \\
	\hline
  \end{tabular}
  \end{threeparttable}
  \end{adjustbox}
\end{table}


\bibliographystyle{apalike}
\bibliography{reference}

\end{document}